\makeatletter
\documentclass[preprint,groupedaddress,showpacs,notitlepage,nofootinbib]{revtex4-1}
\usepackage{epsfig,graphicx,times}
\usepackage{amstext}
\usepackage{amsmath}            
\usepackage{amssymb}            
\usepackage{graphicx}           
\usepackage{latexsym}
\usepackage{hyperref}
\usepackage{bm}
\usepackage{color}

\begin{document}
\title{Resonance Fluorescence from a Two-level Artificial Atom Strongly Coupled to a Single-mode Cavity}
\author{Z.H. Peng}\email{zhihui.peng@riken.jp}
\affiliation{Key Laboratory of Low-Dimensional Quantum Structures
and Quantum Control of Ministry of Education, Department of Physics
and Synergetic Innovation Center for Quantum Effects and
Applications, Hunan Normal University, Changsha 410081, China}
\affiliation{Center for Emergent Matter Science, RIKEN, Wako, Saitama 351-0198, Japan}
\author{D. He}
\affiliation{Key Laboratory of Low-Dimensional Quantum Structures
and Quantum Control of Ministry of Education, Department of Physics
and Synergetic Innovation Center for Quantum Effects and
Applications, Hunan Normal University, Changsha 410081, China}
\author{Y. Zhou}\affiliation{Center for Emergent Matter Science, RIKEN, Wako, Saitama 351-0198, Japan}
\author{J.H. Ding}
\affiliation{ School of integrated circuits,Tsinghua University,
Beijing 100084, China}
\author{J. Lu}\affiliation{Key Laboratory of Low-Dimensional Quantum Structures
and Quantum Control of Ministry of Education, Department of Physics
and Synergetic Innovation Center for Quantum Effects and
Applications, Hunan Normal University, Changsha 410081, China}
\author{L. Zhou}\affiliation{Key Laboratory of Low-Dimensional Quantum Structures
and Quantum Control of Ministry of Education, Department of Physics
and Synergetic Innovation Center for Quantum Effects and
Applications, Hunan Normal University, Changsha 410081, China}
\author{Jie-Qiao Liao}\affiliation{Key Laboratory of Low-Dimensional Quantum Structures
and Quantum Control of Ministry of Education, Department of Physics
and Synergetic Innovation Center for Quantum Effects and
Applications, Hunan Normal University, Changsha 410081, China}
\author{L.M. Kuang}\affiliation{Key Laboratory of Low-Dimensional Quantum Structures
and Quantum Control of Ministry of Education, Department of Physics
and Synergetic Innovation Center for Quantum Effects and
Applications, Hunan Normal University, Changsha 410081, China}
\affiliation{Synergetic Innovation Academy for Quantum Science and Technology, Zhengzhou University of Light Industry, Zhengzhou 450002, China}
\author{Yu-xi Liu}\email{yuxiliu@mail.tsinghua.edu.cn}
\affiliation{ School of integrated circuits,Tsinghua University,
Beijing 100084, China}
\affiliation{Beijing National Research Center for Information Science and Technology (BNRist),
Beijing 100084, China}
\author{Oleg V. Astafiev}
\email{oleg.astafiev@rhul.ac.uk}
\affiliation{Skolkovo Institute of Science and Technology, Nobel str. 3, Moscow, 143026, Russia}
\affiliation{Moscow Institute of Physics and Technology, Institutskiy Pereulok 9, Dolgoprudny 141701, Russia}
\affiliation{Royal Holloway, University of London, Egham Surrey TW20 0EX, United Kingdom}
\affiliation{National Physical Laboratory, Teddington, TW11 0LW, United Kingdom}
\author{J.S. Tsai}
\affiliation{Department of Physics, Tokyo University of Science, Kagurazaka, Tokyo 162-8601, Japan}
\affiliation{Center for Emergent Matter Science, RIKEN, Wako, Saitama 351-0198, Japan}

\begin{abstract}
We experimentally demonstrate the resonance fluorescence of a two-level artificial atom strongly coupled to a single-mode cavity field. The effect was theoretically predicted thirty years ago by Savage [Phys. Rev. Lett. 63, 1376 (1989)]. The system consists of a superconducting qubit circuit and a one-dimensional transmission line resonator. In addition, a one-dimensional transmission line strongly coupled to the atom serves as an open space. The effect takes place, when a microwave field is applied to the cavity, which in turn is resonantly coupled to the atom. The fluorescence spectrum is measured via the emission into the transmission line. We find that the central peak is determined by the atom spontaneous emission to the open space and the widths of side peaks are largely determined by the coherent interaction between the atom and the cavity, that is, the fluorescence spectrum here is very different from that of the Mollow triplet. We also derive analytical form for the spectrum. Our experimental results agree well with theoretical calculations.

\end{abstract}
\maketitle

\textit{Introduction}. The resonance fluorescence of single two-level atoms in open space is not only very fundamental for studying the light-matter interaction in quantum optics~\cite{Scully-book}, but also very important for quantum information processing, e.g., measurement of quantum states in a nondemolition way~\cite{Yilmaz2010}, detection of quantum coherence~\cite{He2015,Lu2021}, generation of non-classical photon states~\cite{Diedrich1987}, creation of quantum network nodes~\cite{Duan2010}. Mollow showed that the spectrum of the fluorescence, from a single-atom strongly driven by a classical field in resonance, has a triplet structure~\cite{Mollow1969}. The resonance fluorescence was observed in atomic systems (e.g., in Ref.~\cite{Schuda1974}) and then in semiconducting quantum dots (e.g., in Refs.~\cite{Xu2007,Muller2008,Vamivakas2009}). With the rapid development of superconducting quantum circuits~\cite{Gu2018,Kockum2019}, the resonance fluorescence has also been broadly studied via superconducting artificial atoms~\cite{Lu2021,Astafiev2010,Baur2009,Abdumalikov2011,Hoi2012,Campagne-Ibarcq2014,Campagne-Ibarcq2016,Lang2011,Wen2018}. Moreover, the resonance fluorescence could be drastically modified by engineering environment. For example, the resonance fluorescence in a squeezed vacuum~\cite{Carmichael1987} has been observed with a superconducting artificial atom recently~\cite{Toyli2016}.

On the other side, the interaction between a two-level atom and a single-mode cavity is a fundamental system in quantum optics. Therefore, it has attracted attention from theoreticians to the resonance fluorescence of the atoms in a cavity,  and many interesting phenomena were found, e.g., the cavity can enhance or suppress the fluorescence and results in asymmetry of the Mollow triplet~\cite{Lewenstein1987,Quang1993}. The dynamics of the system is affected by coherent processes of the multi-photon atom-cavity interaction, which should be revealed by the modifications of widths of the sidebands in the Mollow triplet. They are determined by the coupling strength between the cavity field and the atom in the strong coupling regime~\cite{Savage1989,Savage2007}, each component of the Mollow triplet can be split into a multiplet~\cite{Freedhoff1993} when the cavity field is strongly coupled to the atoms. However, to the best of our knowledge, all these predictions have not yet demonstrated experimentally because difficulties in the broadband fluorescence spectrum from natural atoms coupled to the cavity. This is because of limited measurement bandwidth through the cavity or a weak emission signal induced by atoms weakly coupled to the open space in normal case.

\begin{figure}
\includegraphics[scale=0.4]{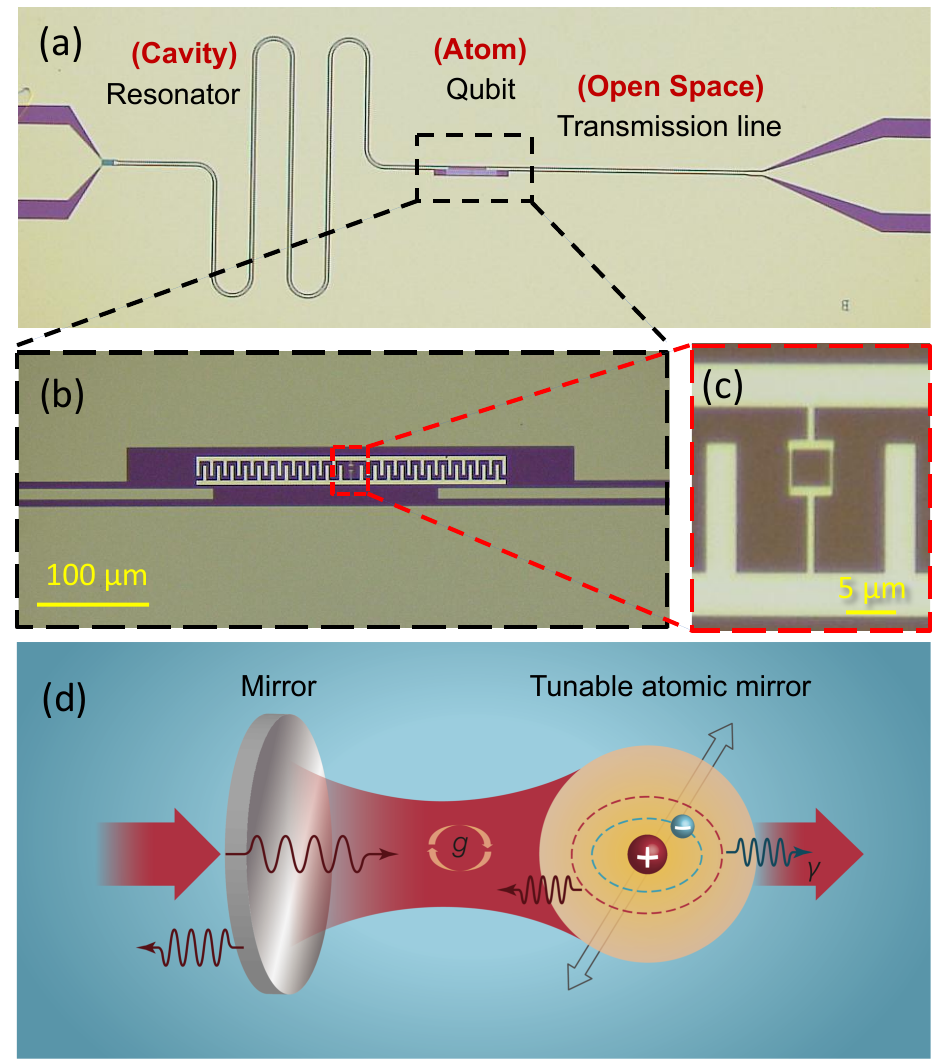}
\caption{(a) Micrograph of the device. A superconducting artificial atom (transmon) is capacitively coupled to a 1D transmission line resonator with meandering structure. Also, the transmon is coupled to a 1D transmission line (on the left side) used to directly measure the reflection spectrum from the atom. (b) Magnified micrograph of the transmon. (c) Magnified part of the SQUID loop of the transmon. (d) Sketch of an atom simultaneously strongly coupled to a cavity and an open space. The transmon can be considered as a part of the cavity and acts as the cavity boundary, that is, it works as a tunable atomic mirror of the cavity. The atom-resonator interaction results in the strong modification of the emission spectrum.}
\label{device}
\end{figure}

In this letter, we show an experimental realization of the resonance fluorescence from the two-level atom strongly coupled to a single-mode cavity field. The effect was predicted in Ref.~\cite{Savage1989} $30$ years ago. In our experiment, the two-level atom and the single-mode cavity are formed by a superconducting qubit circuit and a 1D (one-dimensional) transmission line resonator, respectively. And the difficulties of the limited measurement bandwidth through a cavity or weak emission signal induced by a natural atom weakly coupled to the open space can be overcome by strong coupling of a superconducting two-level atom to both a single-mode field and a 1D transmission line acting as the open space~\cite{Peng2018}. Next, the broadband resonance fluorescence spectrum is directly measured through the 1D transmission line (not through the cavity).

\textit{Experimental setup and basic parameters}. As shown in Figs.~\ref{device}(a)-(c) and schematic diagram in Fig.~\ref{device} (d), our experimental setup is similar to the one previously used in Ref.~\cite{Peng2018} in which a two-level artificial atom is formed by a gap tunable flux qubit circuit. However here, the two-level artificial atom is formed by a superconducting transmon circuit with a SQUID loop~\cite{Koch2007}. Importantly, the atom is strongly coupled to both a one-dimensional coplanar waveguide resonator (CPWR) and a half open 1D transmission line, with an impedance $Z_{0}=50\,\Omega$.
The radiation inside the cavity is reflected from the boundary with the atom and the atom can be tuned to the resonance with the cavity. Therefore, it works as a tunable mirror~\cite{Zhou2008}. Hereafter, we call this two-level artificial atom as the transmon qubit.
\begin{figure}
\center
\includegraphics[scale=0.5]{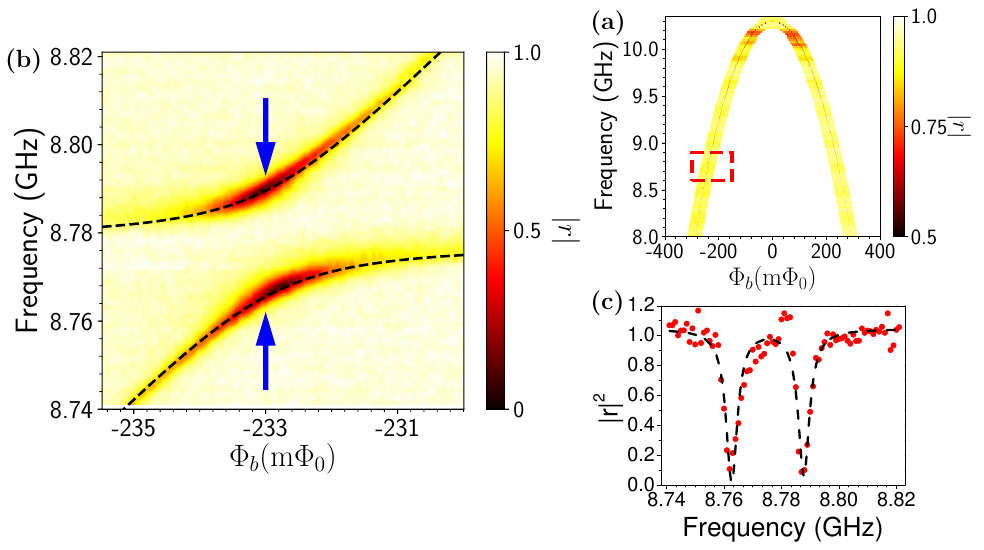}
\caption{(a) Spectroscopy of the artificial atom, measured through a transmission line. Amplitude reflection coefficient $|r|$ as a function of flux bias $\Phi_b$. When the probe signal is in resonance with the transmon qubit, a dip of $|r|$ reveals a dark line. The dashed red rectangle marks the position of the transmon resonantly coupled to the cavity. (b) The anticrossing in the spectroscopy of the transmon qubit which is magnified with the dashed red rectangle range in Fig.2a. The dashed curve is theoretical calculations of energy levels. (c) It shows the splitting at the positions indicated by arrows in (a).}
\label{RefSpec}
\end{figure}

We first characterize basic properties of the device by carrying out the measurement on the device in a dilution refrigerator at a base temperature of about 30 mK. The frequency $\omega_r/2\pi=8.778$ GHz for the fundamental-mode of the CPWR with the decay rate $\kappa/2\pi=5.2\,$ MHz measured by fitting with the Lorentzian curve of a reflection spectrum of the CPWR at a low probing power when the qubit is far detuned from the resonator mode. As shown in Fig.~\ref{RefSpec}(a), the reflection spectrum of the transmon qubit through the transmission line is measured as a function of magnetic flux $\Phi_{b}$, which is applied to the SQUID loop of the transmon qubit, in the frequency domain from $8$ GHz to $10.3\,$GHz with a vector network analyzer. From the data, we extract the maximum Josephson energy $E_{\rm J, max}/h=29.2\,$ GHz and the charging energy $E_{\rm C}/h=0.5\,$ GHz for the transmon qubit, where $E_{\rm C}=e^2/2C_{q}$ with an effective qubit capacitance $C_{q}=39\,$fF. When the biased flux $\Phi_{b}$ is about $-233\,\rm m\Phi_{0}$ with the flux quantum $\Phi_{0}=h/2e$, as shown in Fig.~\ref{RefSpec}(b), the transmon qubit is resonant with the fundamental-mode of the CPWR, and we obtain the coupling strength $g/2\pi=12.0\,$MHz between the transmon qubit and the fundamental mode from fitting of the anticrossing. The relaxation rate of the transmon qubit at the resonant point is $\Gamma_1/2\pi=4.8\,$ MHz. The transmon qubit is strongly coupled to the 1D transmission line, which can be experimentally confirmed by measuring the reflection spectrum of the transmon through the 1D transmission line with a low probing power~\cite{Peng2016}. Furthermore, it is determined that around 60\% of power (quantum efficiency) can be emitted from the excited transmon into the transmission line\cite{Lu2021,Peng2016}. We think that quantum efficiency is limited by the non-radiative decay and 1/f flux noise from two-level systems (TLSs) in the environment \cite{Lu2021c}. As $g > \Gamma_{1}, \kappa$, our system is in the strong coupling regime.

\textit{Resonance fluorescence of a strongly coupled transmon qubit and a single-mode cavity}. To study the resonance fluorescence from the transmon qubit strongly coupled to the fundamental mode of the CPWR in resonance ($\omega_a=\omega_r$), we set the flux bias to $\Phi_{b}=-233\; \rm m\Phi_{0}$ and apply a resonant microwave driving field through the CPWR side at the fundamental-mode frequency $\omega_r$. The full Hamiltonian of the driven system is

\begin{equation}\label{ResonantHaml}
  H=\hbar\omega_{r}a^{\dagger}a+\hbar\frac{\omega_{a}}{2}\sigma_{z}+\hbar g\sigma_{x}(a^{\dagger}+a)+\hbar\Omega\cos(\omega_{r}t)(a^{\dagger}+a).
\end{equation}

Here, $a^{\dagger}$ and $a$ are the creation and annihilation operators of the fundamental mode of the CPWR, respectively, while the Pauli matrices are defined as $\sigma_z=|0\rangle\langle 0|-|1\rangle\langle 1|$ and $\sigma_x=|1\rangle\langle 0|+|0\rangle\langle 1|$ with the ground $|0\rangle$ and excited $|1\rangle$ states of the transmon qubit, the coupling constant $\Omega$ between the driving field and the fundamental mode is proportional to the amplitude of the driving field. We consider the resonant conditions, when $\omega_a = \omega_r$.
\begin{figure}
\center
\includegraphics[scale=0.4]{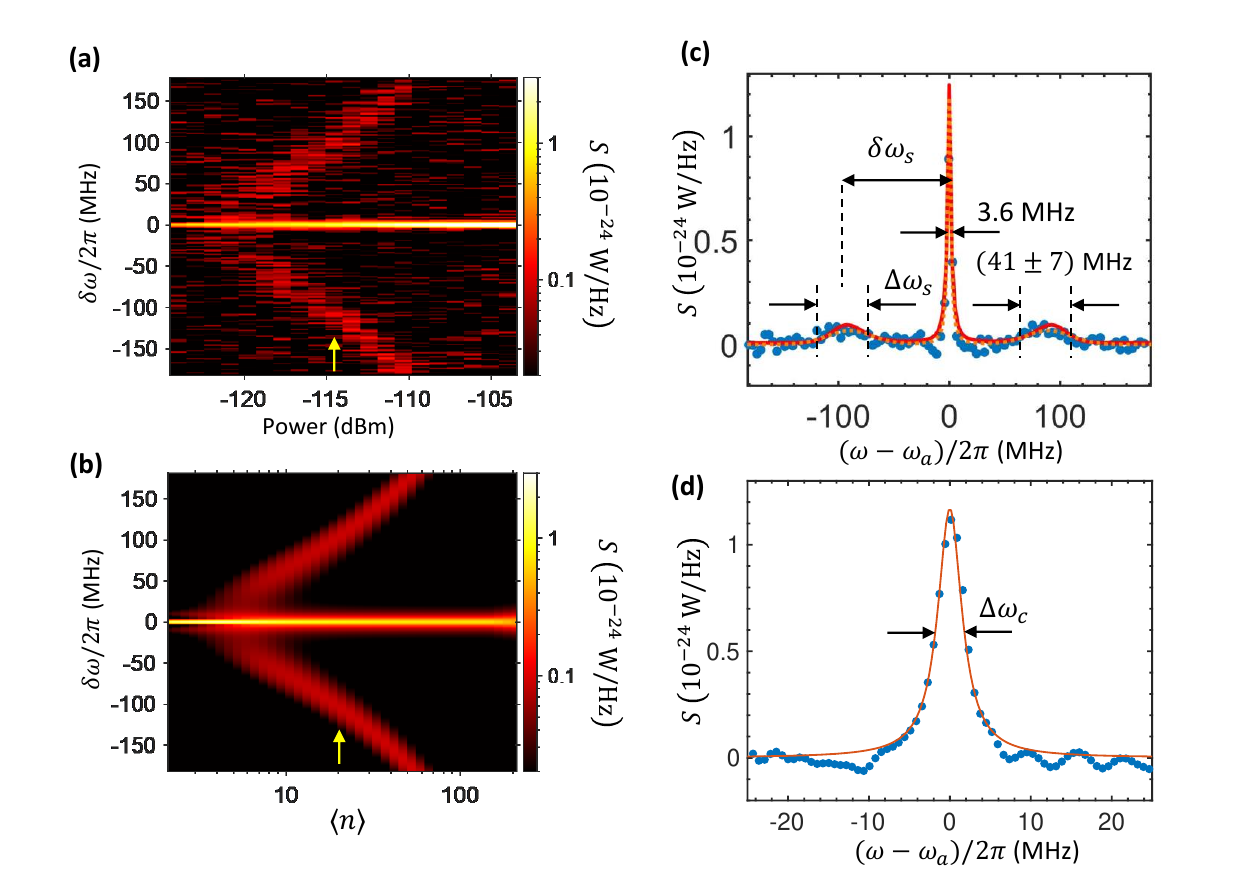}
\caption{The resonance fluorescence of a two-level artificial atom (transmon qubit) strongly coupled to a fundamental mode of the CPWR at the resonance $\omega_a=\omega_r$. The yellow arrow mark the driving power $P=-114.5\,$dBm which corresponds to the Rabi frequency $\Omega/2\pi\approx22.5\,$MHz. (a) The emission spectrum ($S(f) = 2\pi S(\omega)$) as a function of driving power. (b) Theoretical calculations for the emission spectrum with extracted experimental parameters: $\kappa/2\pi=5.2\,$MHz, $\Gamma_1/2\pi=4.8\,$MHz, $g/2\pi=12\,$MHz. (c) An emission spectrum at the driving power $P=-114.5\,$dBm corresponding to the average photon number inside the cavity $\langle n\rangle\sim20.2$. The red curve is the simulation extracted from (b). Fitting with Lorentzian curves of the central peak and by Gaussians the side peaks, the side peaks have widths $\Delta\omega_s/2\pi=(41\pm 7)\,$MHz (orange dashed curve). (d) The blown up central peak from (c) and the effective measured linewidth of the central peak is $\Delta\omega_c/2\pi=3.6\,$MHz.}
\label{Emiss2D}
\end{figure}

The emission spectrum of the strongly coupled transmon qubit coupled to the single-mode cavity is directly measured through the 1D transmission line, when the driving field is resonantly applied to the coupled system through the CPWR side. In this case, the measurement bandwidth through the transmission line is much larger than the measurement bandwidth through the cavity limited by $\kappa$. In the resonance fluorescence measurement, the strong coupling between the transmon qubit and the 1D transmission line results in the efficient collection of emission (almost all power due to atomic relaxation is emitted into the line) and close to the highest signal-to-noise ratio. The measurement setup and the noise background elimination method are similar to those in Ref.~\cite{Astafiev2010}. Figure~\ref{Emiss2D}(a) shows an experimental data of the resonance fluorescence spectra for different driving powers varying from $-123.5$ dBm to $-103.5$ dBm with a step of $1\,$dBm. It shows that the distance between two peaks of the sidebands becomes larger with the increase of the driving power, and the width of the sideband peaks become larger than the width of the resonant peak.

To further understand the mechanism of the resonance fluorescence spectra $S(\omega)$ for the coupled system, we first perform numerical calculations for the correlation function $\langle \sigma_{01}(t+\tau)\sigma_{10}(t)\rangle$  in the steady state limit via the quantum regression theory~\cite{Savage1989} using the master equation $\dot{\rho}=
-i/\hbar [H,\rho]+(\Gamma_1/2)D(\sigma_{-})\rho+\kappa D(a)\rho$ with  $D(A)\rho=2A\rho A^{\dagger}-A^{\dagger}A\rho- \rho A^{\dagger}A$
for the Hamiltonian $H$ in Eq.~(\ref{ResonantHaml}). Then the stationary spectra $S(\omega)$ are obtained via the Fourier transform of the correlation function $\langle \sigma_{01}(t+\tau)\sigma_{10}(t)\rangle$ as ~\cite{Savage1989,Freedhoff1993}

\begin{equation}
S(\omega)=\frac{\hbar\omega_a\Gamma_1}{2\pi}\lim_{t\rightarrow\infty}{\rm Re}\int_{0}^{\infty}\langle \sigma_{01}(t+\tau)\sigma_{10}(t)\rangle\exp(-i\omega \tau)d\tau.
\end{equation}
Numerical simulations with the experimental parameters are shown in Fig.~\ref{Emiss2D}(b), which agrees well with the experimental results.

The emission spectrum, under the resonant drive with the Rabi frequency $\Omega/2\pi\approx22.5\,$MHz corresponding to the driving power $P=-114.5\,$dBm, is further extracted from Fig.~\ref{Emiss2D}(a) and shown in Fig.~\ref{Emiss2D}(c). The linewidth of the central peak in emission spectrum is around $\Delta\omega_c=2\pi\times3.6\,$MHz by fitting with a Lorentzian curve (red curve in Fig. \ref{Emiss2D}(d)). It is expected that the width of the incoherent peak is limited by $\Gamma_1$, however, additional coherent radiation in the form of a narrow peak in the center should lead to an effective narrowing of the combined peak. This is also confirmed by our simulations (see Supplementary materials).
It is clear that the observed resonance fluorescence triplet is dramatically changed comparing with the usual Mollow triplet, as shown in the theoretical result~\cite{Savage1989,Freedhoff1993}. We show that the central peak is much higher than the two side peaks, their ratio is around $11$, which is very different from the standard Mollow triplet, where it is about $3$ and anomalous side peak broadening is additionally observed. The widths of side peaks fitted by $S_{s}(\omega) \exp[-2(\pm\delta\omega_s/\Delta\omega_s)^2]$ (where $\delta\omega_s$ is the side peak positions in respect to the central peak, $S_{s}(\omega)$ is the peak power density) is about $\Delta\omega_s=2\pi\times(41\pm7)\,$MHz. We confirm that the widths of the side peak are not determined by either atomic relaxation rate $\Gamma_1$ or the cavity decay rate $\kappa$.

Below we provide an explanation of the dynamics, which results in the observed spectrum. First, we would like to point out that physics in a system with an atom strongly coupled to and driven from a resonator changes dramatically compared to dynamics of a bare resonator or an atom each connected to an open space only. Emission from a coherent state in a bare resonator driven by a monochromatic wave represents an infinitely narrow spectral peak due to an elastic scattering. Also in a standard Mollow triplet with the atom coupled to the open space only (without a resonator), all peaks are determined by spontaneous emission and therefore have widths defined by $\Gamma_1$ (we consider the case of negligible pure dephasing): The central peak width is $\Gamma_1$ and the sideband peak widths are $3\Gamma_1/2$.
On contrary, when $g > (\kappa, \Gamma_1)$, the coherent atom-resonator interaction begins to play a major role, which drastically modifies the emission spectrum.

\begin{figure}
\center
\includegraphics[scale=0.42]{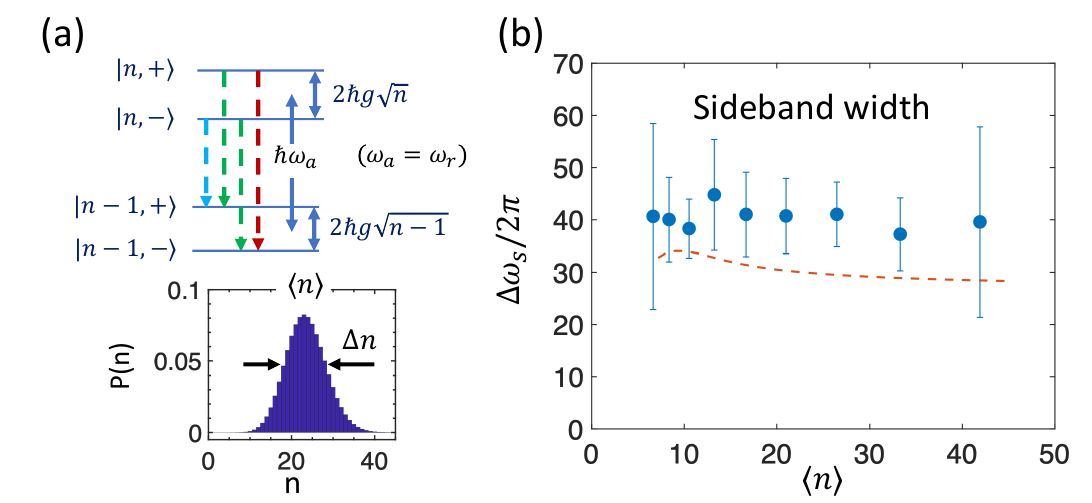}
\caption{(a) Upper panel: Schematic representation of dressed states and all possible relaxations. Low panel: Gaussian photon distribution of the population for the coupled system dressed states with $\langle n\rangle = 20.2$.
(b) The experimentally measured side peak widths from the Gaussian fits. The widths are slightly higher than extracted from simulations (red dashed curve), which in turn is close to $2g$.}
\label{Widths}
\end{figure}

The main features of the emission spectrum presented in Fig.~\ref{Emiss2D}(c) can be explained in the picture of dressed-states 
\begin{eqnarray}\label{rcl}
\begin{array}{rcl}
|n,+\rangle = \frac{|n\rangle|0\rangle + |n-1\rangle|1\rangle}{\sqrt2},  |n,-\rangle = \frac{|n\rangle|0\rangle - |n-1\rangle|1\rangle}{\sqrt2},
\end{array}
\end{eqnarray}
with eigenvalues $E_{|n,+\rangle} = n\hbar\omega_a+\hbar g\sqrt n$ and $E_{|n,-\rangle} = n\hbar\omega_a-\hbar g\sqrt n$  for the atom-CPWR system driven through the resonator field~\cite{Savage1989,Freedhoff1993,Savage2007}.
 Here, $|n\rangle$ is the photon Fock state of the CPWR. There are four spontaneous emission transitions from the initial dressed states $|n,+\rangle$ and $|n,-\rangle$ to the adjacent ones $|n-1,+\rangle$ and $|n-1,-\rangle$ with frequencies $\omega_a\pm g(\sqrt n\pm\sqrt{n-1})$ (a level diagram of Fig.~\ref{Widths}(a)).

The following two transitions $|n,+\rangle \rightarrow |n-1,+\rangle$ and $|n,-\rangle \rightarrow |n-1,-\rangle$ are denoted by dashed green lines in Fig.~\ref{Widths}(a) (see also Supplementary Materials). The corresponding transition frequencies are
$(E_{n,\pm} - E_{n-1,\pm})/\hbar = \omega_a \pm g(\sqrt n - \sqrt{n-1})$. When $g/\sqrt n \leq \Gamma_1$ (in our experiment this is satisfied already with a very few photons), these transitions result in a central emission peak at $\omega_a$ with the width $\Gamma_1$ determined by the atom spontaneous emission to the line and the power density in maximum is
$\sim\hbar\omega_a$. Therefore the central peak with large mean photon number $\langle n\rangle$ is similar to the central peak of the standard Mollow triplet and the power spectral density can be written as

\begin{equation}
S_c(\delta\omega_c) = \frac{1}{2\pi}\frac{\hbar\omega_a \Gamma_1}{4} \frac{\Gamma_1} {\delta\omega_c^2 + (\Gamma_1/2)^2},
\label{Sc}
\end{equation}
where $\delta\omega_c = \omega - \omega_a$. The factor $1/4$ is a result of two squared transition matrix elements ($2\times 1/4$) due to two processes (green arrows in Fig.~\ref{Widths}(a)) contributing to the central peak and another factor 1/2 comes from the half population of each level in the stationary conditions.

The situation is very different for the sidebands. Two other transitions $|n,+\rangle \rightarrow |n-1,-\rangle$ and $|n,-\rangle \rightarrow |n-1,+\rangle$ (blue and red dashed lines in Fig.~\ref{Widths}(a)) result in the appearance of sidebands with frequencies $(E_{n,+} - E_{n-1,-})/\hbar = \omega_a + g(\sqrt n + \sqrt{n-1})$ and $(E_{n,-} - E_{n-1,+})/\hbar = \omega_a - g(\sqrt n + \sqrt{n-1})$. We can estimate the peak properties when the mean photon number in the resonator $\langle n\rangle$ is large ($\langle n\rangle \gg 1$). The Gaussian photon distribution in the coherent state is $P(n) = (2\pi \langle n\rangle)^{-1/2}\exp[-\delta n^2/2\langle n\rangle]$, where $\delta n = n - \langle n\rangle$ and the peak width $\Delta n \approx 2\sqrt{\langle n\rangle} $ (see Fig.~\ref{Widths}(a)).
The transition frequencies can be approximated by $\omega_a \pm (\omega_s + \delta\omega_{sn})$, where
$\omega_s = 2g\sqrt{\langle n\rangle}$, $\delta\omega_{sn} = g\delta n/\sqrt{\langle n\rangle})$.
The spectral density can be calculated as $S_s(\delta\omega_s) =  (2\pi/8)\hbar\omega_a\Gamma_1 P(\omega_{s})$, where $P(\delta\omega_s)d\delta\omega_s = P(n)dn$, and we arrive to

\begin{equation}
S_s(\delta\omega_s) =  \frac{1}{2\pi}\frac{\hbar\omega_a\Gamma_1}{8} \frac{\sqrt{2\pi}}{g} \exp\Big[-\frac{1}{2}\Big(\frac{\delta\omega_s}{g}\Big)^2\Big] ,
\label{Ss}
\end{equation}
where $\delta\omega_s = \omega - \omega_s$ (see Supplementary materials). The prefactor 1/8 is twice less than in Eq.~(\ref{Sc}) because only one process (either blue or red in Fig.~\ref{Widths}(a)) contributes to each side peak. The side peak widths asymptotically goes to $2g$ at $\langle n\rangle \rightarrow\infty$. Note that the side peak widths are larger than $2g$ when the driving field is not approaching the limit of $\Omega\gg g$ like the results presented in Fig.~\ref{Emiss2D}c.

In our experiment, the resonator is driven into the coherent state with the mean photon number inside the cavity $\langle n\rangle=P/\hbar\omega_r\kappa'\sim20.2$, where $\kappa' = (\kappa + \Gamma_1)/2$ is an effective decay rate of photons via the dressed states at the conditions marked by an yellow arrow in Fig.~\ref{Emiss2D}(a), corresponding to Fig.~\ref{Emiss2D}(c). The distance between the sideband peaks and the central peak, $\delta\omega_s$ is expected to be $2g\sqrt{n}\sim 2\pi\times 108\,$MHz, which is close to the experimental value $2\pi\times 90\,$MHz. In the limit of $\Omega\gg g$, the artificial atom will slightly perturb the photon number statistics inside the resonator. Therefore, as shown in the inset plot of Fig.~\ref{Widths}(a), the population distribution of the dressed states can be approximated with the coherent-state Gaussian distribution. The peak maximum of the distribution is at $n=(\Omega/\kappa')^2=20.2$ and width at one standard deviation is $\sqrt{n}\sim4.5$.
The width of the sidebands from simulations is $\Delta\omega_s^{\prime}/2\pi=30\,$MHz which is very close to $2g$ and is reasonably well agreed with the experimental value of $\Delta\omega_s$. Integration of the powers in Fig.~\ref{Emiss2D}(c) give ratio of central to two-side-peaks 1.1 supporting the proposed mechanisms.

Experimentally measured side peak widths versus mean photon number $\langle n\rangle$ are shown in Fig.~\ref{Widths}(b), where $\langle n\rangle$ is calculated from the maxima of the side peaks $\delta\omega_s$ in respect to the central peak frequency $\omega_c$ according to $\langle n\rangle = (\delta\omega_s/2g)^2$. Additional broadening can be attributed to the dephasing effect from the atom and the deviation of the state inside the cavity from the classical coherent state due to scattering on the atom.

In conclusion, we study the resonance fluorescence from a superconducting artificial atom strongly coupling to a cavity. The dynamics of the system and the fluorescence spectra are drastically changed when the cavity is driven in contrast to the standard Mollow triplet. The widths of the sidebands are determined by coherent coupling strength between the artificial atom and the cavity field. Our system can also be used to demonstrate other resonance fluorescence phenomena of the atoms in a cavity, e.g. asymmetry of the Mollow triplet~\cite{Lewenstein1987,Quang1993}. All the results demonstrate that the superconducting artificial atom is an ideal platform for quantum optics and quantum physics research.

\begin{acknowledgements}
This work was funded by NSFC under Grant No.~12074117, No.~61833010, No.~12061131011, No.~11935006. This work was supported by JST [Moonshot R$\&$D][Grant No.~JPMJMS2067], the New Energy and Industrial Technology Development Organization (NEDO) under Grant No.~JPNP16007, and CREST, JST. (Grant No. JPMJCR1676). JL is supported by the NSFC under Grant No.~12075082. LZ is supported by the NSFC under Grant Nos.~11975095. J.-Q.L. was supported in part by NSFC (Grant No. 12175061), and Hunan Science and Technology Plan Project (Grants No. 2017XK2018, No. 2021RC4029, and 2020RC4047). LMK is supported by the NSFC under Grant Nos.~1217050862, and 11775075. YXL is supported by the Key R\&D Program of Guangdong province under Grant No.~2018B030326001 and NSFC under Grant No.~11874037. OVA was funded by the Russian Science Foundation under Project No. 21-42-00025.
\end{acknowledgements}
\bibliography{ResFluorRefs}
\end{document}